\documentstyle[preprint,aps]{revtex}
\begin{document}
\title{Ab initio Molecular Dynamics in Adaptive Coordinates}

\author{Fran\c{c}ois Gygi}

\address{
 Institut Romand de Recherche Num\'erique
 en Physique des Mat\'eriaux (IRRMA)\\
 CH-1015 Lausanne, Switzerland
}

\maketitle
\begin{abstract}
 We present a new formulation of {\em ab initio}
molecular dynamics which exploits the efficiency of plane waves in
adaptive curvilinear coordinates, and thus provides an accurate
treatment of first-row elements.
 The method is used to perform a molecular dynamics simulation of
the CO$_2$ molecule, and allows to
reproduce detailed features of its vibrational spectrum such as
the splitting of the Raman $\sigma^+_g$ mode caused by Fermi resonance.
 This new approach opens the way to highly accurate
{\em ab initio} simulations of organic compounds.
\end{abstract}

\pacs{71.10.+x, 33.10.-n, 31.15.+q, 02.70.Ns}

 An increasing number of applications of {\em ab initio} molecular
dynamics (MD) has given considerable insight into
dynamical and statistical properties of a wide range of materials \cite{CaPa}.
 However, the application of this method to important fields such as
organic chemistry, or to the study of oxide materials, has been hindered by
difficulties in representing accurately first row elements using
plane waves and pseudopotentials.
 Implementations of {\em ab initio} MD based on plane waves
are intrinsically limited by the slow convergence of the Fourier
expansions of wavefunctions and potentials.
 This limitation is particularly severe in the case of first row elements
for which both wavefunctions and potentials are rapidly varying.
 Although some progress has been made recently to circumvent this problem
by introducing modified pseudopotentials for first-row elements
\cite{TrMa,Van},
a unified approach treating all atomic species on the same footing, and
insuring the same accuracy and transferability of all potentials in various
environments is still lacking.
 A method which relies on well established norm-conserving pseudopotentials
\cite{Hamann1},
and does not necessitate special approximations for the representation of first
row elements is essential e.g.\ for the study of materials under high
pressure, or for MD simulations carried out at high
temperatures. In such situations, interatomic distances can become
substantially smaller than equilibrium distances, which requires a high
transferability of pseudopotentials.
 Furthermore, it would be highly desirable to take advantage of the large
body of information accumulated over the years about
conventional norm-conserving pseudopotentials.

 In this paper, we present a new formulation of {\em ab initio}
MD, which exploits the efficiency of plane waves in adaptive curvilinear
coordinates, and therefore allows for an accurate treatment of all elements
using conventional norm-conserving pseudopotentials.
 An application to the calculation of vibrational properties of the CO$_2$
molecule yields results comparable to those obtained with the most elaborate
quantum chemistry methods, but also allows to observe anharmonic effects
on its vibrational spectrum.

 Our approach is based on the
method of plane waves in adaptive coordinates which was recently proposed
\cite{Gy1,Gy2}
to perform highly accurate electronic structure calcualtions,
within the density functional, pseudopotential formalism.
 This method has been succesfully applied to the determination of
the equilibrium configuration of small molecules containing first-row
elements \cite{Gy2} and to selected phases of SiO$_2$ \cite{Hamann}.
 The basis set used in this method consists of
plane waves in curvilinear coordinates
\begin{equation}
 \chi_{\bf k}({\bf x}) =
  \frac{1}{\sqrt{\Omega}}
 \left| \frac{\partial {\xi}}{ \partial x} \right|^\frac{1}{2}
 \exp [i{\bf k}{\xi}({\bf x})]
 \label{eq:chidef}
\end{equation}
where
$|\partial {\xi}/\partial x|$ is the determinant of the Jacobian
of the coordinate transformation
${\bf x} \mapsto {\xi}({\bf x})$,  and
${\bf k}$ is a reciprocal lattice vector.
 The functions $\chi_{\bf k}({\bf x})$ reduce
to ordinary plane waves in the
limit where ${\bf x}(\xi) = \xi$, i.e.\ when the coordinates are Euclidean.
 They form a complete and orthonormal basis for {\em any} choice of
coordinates.
 In the approach of Ref.~\cite{Gy2}, coordinates are allowed to adapt freely
so as to minimize the total energy.
 This implies that the basis set does not depend
explicitly on ionic positions, so that no Pulay forces
on ions have to be calculated.
 It also allows for an optimal
adaptation of the basis set, and therefore leads to a rapid convergence of the
calculation as the number of basis functions is increased.
 In its original formulation, the method of adaptive coordinates
could be applied to {\em ab initio} MD simulations
by either i) reoptimizing the coordinates and the electronic
wavefunction coefficients at every ionic step, or ii) assigning a fictitious
mass to the coordinates and solving a second order equation of motion
for the electronic, ionic and coordinate degrees of freedom.
 The first approach is workable \cite{Gy3}, although in practice a very
accurate minimization of electronic and coordinate degrees of freedom must
be performed at each ionic step in order to avoid the accumulation of
systematic errors.
 The second approach, although it can be used succesfully in a simulated
annealing search for minimum energy configurations, is not applicable for
MD simulations since a large fictitious mass must be assigned to
the coordinates in order to keep their dynamical evolution stable.
 When combined with usual integration time steps (5-10 a.u.) this large
fictitious mass causes a sizeable energy transfer to occur between the
ionic and coordinate degrees of freedom during the MD simulation.
As a consequence, the potential energy surface followed by the ions
does not correspond to the Born-Oppenheimer surface,
and the resulting ionic trajectories are meaningless. This energy transfer
between degrees of freedom could be avoided by using very small time steps,
although this leads in practice to an unacceptably large number of steps.

 Another approach consists in noting that
efficient basis sets can also be defined by choosing {\em explicitly} the
coordinate transformation $\xi \mapsto {\bf x}(\xi)$.
 In what follows, we define ${\bf x}(\xi)$ in such a way that
the effective plane wave energy cutoff is increased in a prescribed
way near atomic positions.
 The coordinate transformation
${\bf x} \mapsto {\xi}({\bf x})$ is defined by specifying its {\em inverse},
i.e.\
\begin{equation}
  \xi^i = x^i - \sum_\alpha f^i_\alpha({\bf x} - {\bf R}_\alpha)
  \label{eq:xidef}
\end{equation}
where ${\bf R}_\alpha$ is the position of atom $\alpha$, and $f^i_\alpha$
describes the deformation of coordinates around atom $\alpha$ in the
direction $i$.
 The form chosen for the functions $f^i_\alpha$ must be
such as to increase the effective energy cutoff on the atomic
site ${\bf R}_\alpha$.
 We use the following definition
\begin{equation}
  f^i_\alpha({\bf x}-{\bf R}_\alpha) =
  - (x^i-R^i_\alpha) f_\alpha(|{\bf x}-{\bf R}_\alpha|)
\end{equation}
where $f_\alpha(r)$ is a positive, rapidly decaying function.
 The definition of $\xi({\bf x})$ is then given by
\begin{equation}
  \xi^i = x^i + \sum_\alpha
  (x^i-R^i_\alpha) f_\alpha(|{\bf x} - {\bf R}_\alpha|).
  \label{eq:nlin}
\end{equation}
 The Jacobian of the coordinate transformation, which is directly related
to the plane wave effective energy cutoff \cite{Gy2}, can be computed directly
from the $f^i_\alpha$'s
\begin{eqnarray}
 \frac{\partial \xi^i}{\partial x^j} =
 \delta_{ij} +
 \sum_\alpha \left[
 \mbox{\rule{0 em}{3 ex}}
 \delta_{ij} f_\alpha(|{\bf x} - {\bf R}_\alpha|)
 \right. & & \nonumber \\
 \left.
 + \frac{(x^i-R^i_\alpha) (x^j-R^j_\alpha)}{|{\bf x} - {\bf R}_\alpha|}
 \left. \frac{\partial f_\alpha(r)}{\partial r}
 \right|_{r=|{\bf x} - {\bf R}_\alpha|}
 \right].
 & &
 \label{eq:jac}
\end{eqnarray}
The Jacobian (\ref{eq:jac}) is symmetric, which implies in particular that the
coordinates are curl free
\[
   \frac{\partial \xi^i}{\partial x^j} -
   \frac{\partial \xi^j}{\partial x^i} = 0.
\]
 Various choices can be made for the definition of
the function $f_\alpha(r)$.
 We use
\begin{equation}
  f_\alpha(r) = A_\alpha \; \mbox{sech} \frac{r}{a_\alpha}
\end{equation}
where $a_\alpha$ is a parameter defining the range
over which the energy cutoff is increased around atom $\alpha$ \cite{falpha}.
 The constant $A_\alpha$ is chosen so as to reproduce a given
maximum energy cutoff $E^\alpha_{\rm cut}$ at the atomic site
and is given by
\begin{equation}
 A_\alpha = \left( \frac{E^\alpha_{\rm cut}}{E^0_{\rm cut}}
 \right)^\frac{1}{2} - 1
\end{equation}
where $E^0_{\rm cut}$ is the plane wave energy cutoff in
Euclidean coordinates.
 The choice of the value of $a_\alpha$ is not critical.
 It can be deduced e.g.\ from a calculation of the electronic
structure of an isolated atom carried out with fully adaptive coordinates.
 The coordinates ${\bf x}(\xi)$, which are needed for the calculation
of the total electronic energy, are then obtained by
solving the nonlinear equation (\ref{eq:nlin}) iteratively
using Newton's method, and by taking advantage of the fact that the Jacobian
matrix (\ref{eq:jac}) is known analytically.
 The solution of Eq.\ (\ref{eq:nlin}) must be repeated whenever
atomic positions are updated during a MD simulation.
 Clearly, the solution obtained for the previous atomic positions provides
a good starting point
for the iterative solution of Eq.(\ref{eq:nlin}), and in practice,
only a few iterations are necessary.
 The basis set defined by these coordinates depends only on
ionic positions and on the functions $f_\alpha$.
 This effectively constrains the regions of enhanced energy cutoff
to follow smoothly the atoms during the course of a MD simulation,
insuring a proper description of the electronic wavefunctions
in the vicinity of atoms.
 Using such an explicit choice of coordinates will in general lead to
a basis set which is less optimal than in the fully adaptive approach.
However, since the basis set is complete for any choice of coordinates,
the accuracy of the results can be
improved systematically by increasing the value of $E^0_{\rm cut}$.

 Once the coordinates ${\bf x}(\xi)$ are defined, all
quantities needed for the calculation
of the energy and of the forces on the electronic degrees of freedom can
be deduced from them, and the procedure is identical to that presented in
Ref.\ \cite{Gy2}.
 Similarly, ionic forces can be calculated following
Ref.\ \cite{Gy2}, except for an additional Pulay term which comes from the
explicit dependence of the basis set on atomic positions.
 The Pulay component of the force on the atomic coordinate $R^s_\alpha$
is given by
\[
 F^s_\alpha =
 - \int \frac{\partial \xi^q}{\partial R^s_\alpha}
 \frac{\delta E}{\delta \xi^q}
 \; d^3\xi
\]
where
\[
  \frac{\delta E}{\delta \xi^q} =
  \frac{\partial x^p}{\partial \xi^q}
  \frac{\delta E}{\delta x^p}
\]
and summation over repeated indices is implied.
 The quantities $\delta E / \delta x^p$ are the forces acting on coordinates,
which are described in detail in Ref.\ \cite{Gy2}.

 This approach was used to calculate the vibrational spectrum
of the CO$_2$ molecule, a prototypical example of small organic molecule.
 In spite of its small size, the calculation of its
vibrational properties including anharmonic effects
already represents a difficult test case for traditional
{\em ab initio} quantum chemistry methods \cite{Dyk}.
 We used norm-conserving, non-local pseudopotentials\cite{BHS},
in their separable form \cite{KlBy}, and the local density approximation
(LDA) for exchange and correlation\cite{CeAl,PeZu}.
 The parameters defining the coordinates were $a_\alpha=1.4$ a.u. and
$A_\alpha$ was chosen so as to produce an effective
plane wave energy cutoff of 90 Ry on oxygen and 40 Ry on carbon,
starting from an initial Euclidean energy cutoff of $E_{\rm cut}^0 = 20$ Ry.
 The molecule was placed in a simulation cell of $18\times14\times14$ a.u.\,
and the ground state electronic structure of CO$_2$ was first determined
in a fixed asymmetric ionic configuration corresponding to a small
distortion away from its equilibrium geometry, and involving all three
modes of deformation of the molecule (bending, symmetric and antisymmetric
stretching).
 The MD simulation was started with an equilibration
run of 0.1 ps, followed by a MD simulation of 0.2 ps during which
ionic trajectories were collected.
 The behaviour of the Kohn-Sham energy, the kinetic energy of ions, the
fictitious kinetic energy of the electrons, and the total energy
is shown in Fig.~1 for a fraction of the total simulation time.
 The absence of any noticeable drift in the total energy confirms that
the equations of motion for both ions and electrons are integrated accurately.
 In order to illustrate a typical coordinate transformation ${\bf x}(\xi)$,
we show in Fig.~2 a snapshot of ${\bf x}(\xi)$ in a plane containing
the molecule.

 The vibrational spectrum of CO$_2$
was calculated from the data collected during the 0.2 ps simulation
using the maximum entropy method \cite{numrec} with 150
poles in the fitting procedure,
and then used to extract vibrational frequencies.
 The experimentally measured low-frequency vibrational spectrum
of CO$_2$ consists
of two infrared active modes at 667 cm$^{-1}$ ($\pi_u$)
and 2349 cm$^{-1}$ ($\sigma^+_u$), and a Raman $\sigma^+_g$ doublet
at 1286 and 1388 cm$^{-1}$ split by Fermi resonance \cite{Herz}.
 This splitting is caused by a strong enhancement of
anharmonic coupling between the $\sigma^+_g$ and $\pi_u$ modes which
originates in the near degeneracy of the overtone of
the $\pi_u$ mode with the fundamental of the $\sigma^+_g$ mode.
 Reproducing this splitting clearly represents a challenge for any
{\em ab initio} method.

 After the first 0.05 ps of MD simulation,
the calculated power spectrum already exhibits
peaks near 630 and 2380 cm$^{-1}$, which can be assigned to the
$\pi_u$ and $\sigma^+_u$ modes respectively.
 Another peak corresponding
to the $\sigma^+_g$ mode appears near 1400 cm$^{-1}$ and is broader,
indicating mixing between the fundamental of the
$\sigma^+_g$ mode and the overtone of the $\pi_u$ mode.
 After completion of the MD simulation, we observe sharp
$\pi_u$ and $\sigma^+_u$ peaks at 624 and 2310 cm$^{-1}$ respectively,
whereas the $\sigma^+_g$ Raman peak splits into
two well resolved peaks at 1356 and 1392 cm$^{-1}$.
 The magnitude of the splitting between the two Raman
peaks is comparable to the experimental splitting.
 However, since the present simulation treats the ions classically,
a more detailed comparison with experiment would be meaningless.
 The calculated center of gravity of the Raman doublet
lies at 1371 cm$^{-1}$, to be compared to an experimental value of
1337 cm$^{-1}$.
 In order to check the convergence of the above results, the full
simulation was repeated starting from an initial energy
cutoff ($E_{\rm cut}^0$) of 30 Ry, which produces a maximum energy
cutoff ($E_{\rm cut}^\alpha$) of 138 Ry on oxygen atoms.
 The resulting frequencies are 642, 1368, 1428, and 2353 cm$^{-1}$ for
the $\pi_u$, the Raman $\sigma_g^+$ doublet and the $\sigma_u^+$ modes
respectively. The agreement with experiment improves for the $\pi_u$
and $\sigma_u^+$ modes, but degrades slightly for the Raman doublet.
 Errors in the calculated frequencies are of the order of 40 cm$^{-1}$ for
the $\pi_u$ and for the higher mode of the Raman doublet,
and 80 cm$^{-1}$ for the lower mode of the Raman doublet---the smaller
error (4 cm$^{-1}$) obtained for the $\sigma_u^+$ mode
being considered fortuitous.
 These results are summarized in Table I. and the power spectrum is shown
in Fig.~3.
 This kind of agreement with experiment is excellent for a
first principles approach.

 Quantum chemistry methods based on energy gradients have
been used to compute the
 {\em harmonic} vibrational frequencies of CO$_2$
including various combinations of basis sets
and treatments of electronic correlations \cite{ThBr}.
 Harmonic frequencies cannot be compared directly to experiment,
but only to frequencies
deduced by fitting a potential energy function to experimental results
\cite{CiCh}.
 Although frequencies obtained at the SCF level are unsatisfactory
---errors can be as large as 150 cm$^{-1}$---the errors
obtained with the coupled-cluster methods (CCSD and CCSD(T)) \cite{Bar}
are smaller than 30 cm$^{-1}$, an accuracy comparable to that
of our results.
 We conclude that, in the case of CO$_2$, the LDA treatment of exchange
and correlation yields excellent results, which are comparable to those of the
most accurate methods of quantum chemistry.
 Furthermore, the ability to perform {\em ab initio} MD allows one to
observe anharmonic behaviour such as the Fermi resonance discussed above.
 MD simulations are beyond the present capabilities of quantum chemistry
methods.

 In conclusion, we have presented a new formulation of {\em ab initio}
molecular dynamics which uses plane waves in adaptive coordinates.
 This approach allows to treat all atoms in the simulation on an equal
footing, using norm-conserving, non-local pseudopotentials.
 The large enhancement of the plane wave effective energy cutoff achieved
near atoms allows one to use conventional pseudopotentials without any
compromise on their transferability.
 A calculation of the vibrational spectrum of CO$_2$
demonstrates the accuracy of the method and its applicability to
{\em ab initio} MD simulations of organic compounds.
 A study of the equilibrium and vibrational properties of
a simple carbohydrate molecule using this approach is under way \cite{Gy4}.

 I would like to thank G.~Galli for discussions, and
useful comments on the manuscript.

\begin{figure}
\caption{
  Evolution of the Kohn-Sham energy (solid line), the ionic kinetic energy
(dahsed line), and the fictitious electronic kinetic energy (dotted line)
during part of the MD simulation. The total energy (dash-dotted line) shows
fluctuations of the order of $10^{-4}$ eV and no noticeable drift.
}
\end{figure}

\begin{figure}
\caption{
  Image of a regularly spaced rectangular grid in $\xi$ space under the
  map ${\bf x}({\xi})$, calculated in a plane containing the CO$_2$ molecule.
  The left and right circles represent oxygen atoms,
  and the center circle represents the carbon atom.
}
\end{figure}

\begin{figure}
\caption{
  Power spectrum of the vibrations of the CO$_2$ molecule extracted from
a trajectory of 0.2 ps. The spliting of the symmetric stretching mode
around 1400 cm$^{-1}$ is clearly visible.
}
\end{figure}

\begin{table}
 \begin{tabular}{lccc}
  & $\pi_u$ & $\sigma_g^+$ & $\sigma_u^+$ \\
 \tableline
   $E_{\rm cut}^0$ = 20 Ry  &  624 &  1356,1386 & 2310 \\
   $E_{\rm cut}^0$ = 30 Ry  &  642 &  1368,1428 & 2353 \\
   Exp. &  667 &  1286,1388 & 2349 \\
 \end{tabular}

\caption{
 Vibrational frequencies of CO$_2$ as obtained
from {\em ab initio} simulations carried out with $E_{\rm cut}^0 =$ 20 Ry and
30 Ry, and from experiment (Exp.).
}
\end{table}


\begin{references}
\bibitem{falpha} Other functions have been tried and yield equally good
results, e.g.\ $f_\alpha(r) =
A_\alpha (a_\alpha/r) \tanh (r/a_\alpha) \exp -(r/b_{\alpha})^2 $,
where $a_\alpha$ is the range of enhancement of the energy cutoff
and $b_\alpha$ determines a distance over which Euclidean coordinates
are recovered.
\end{references}
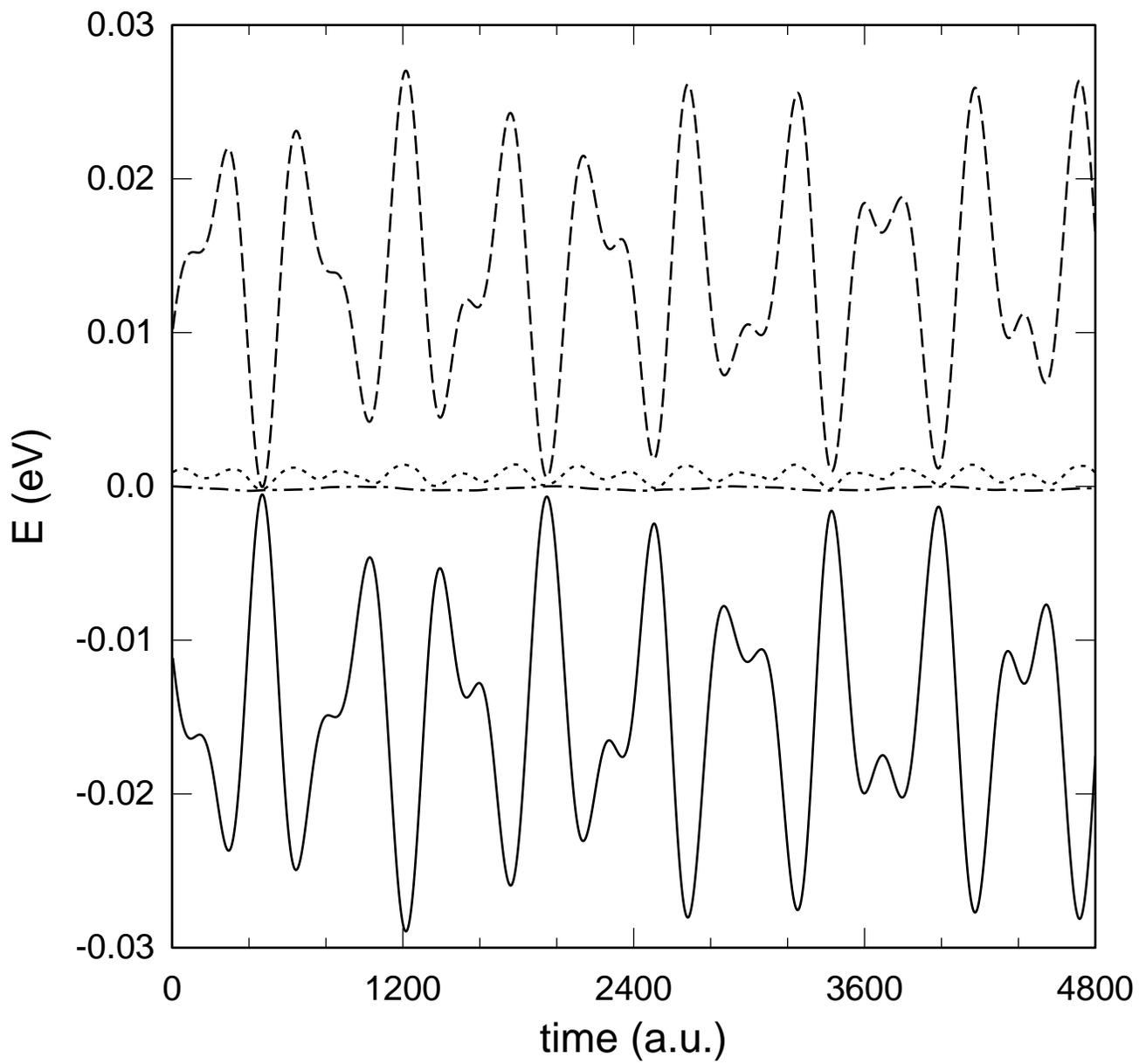
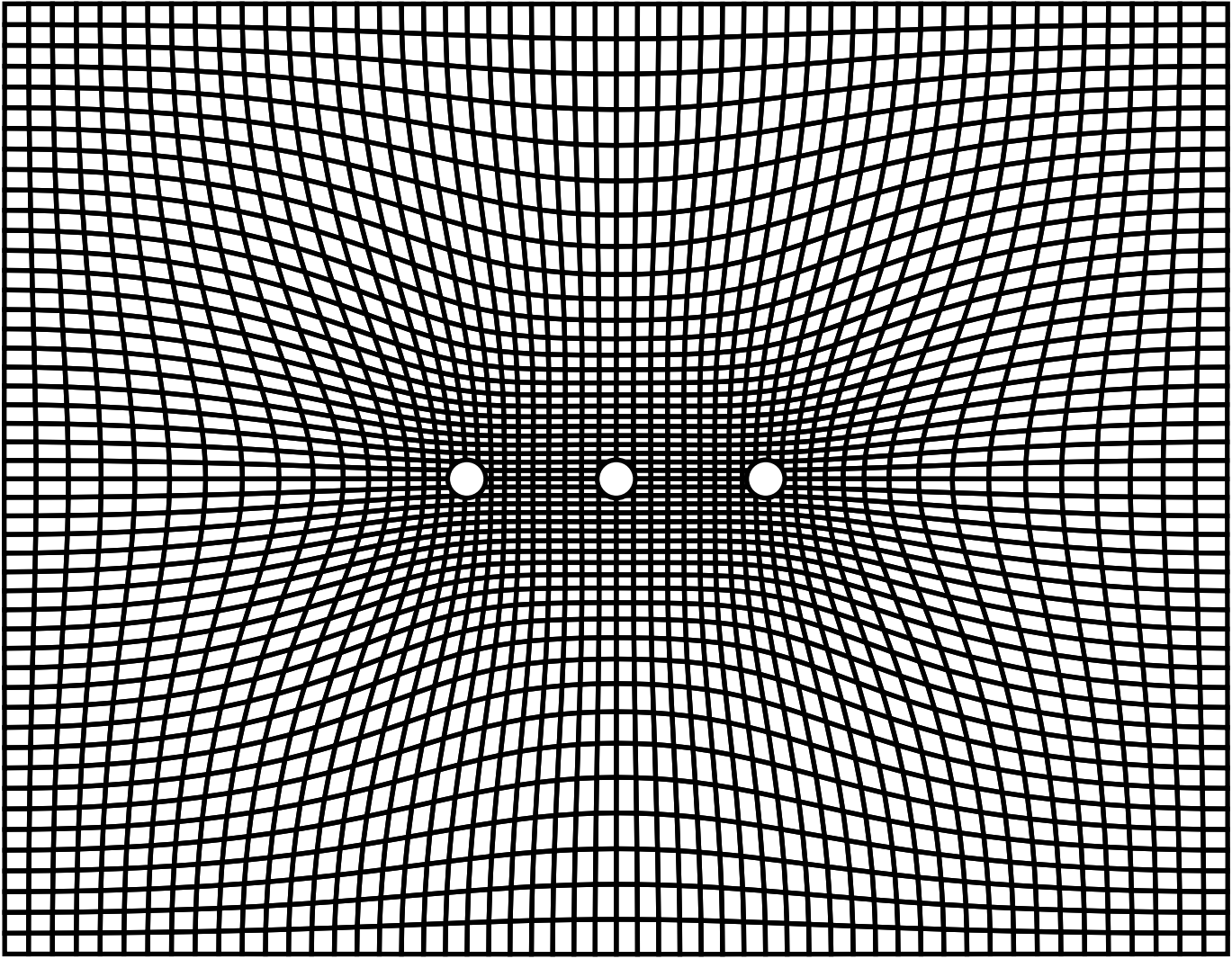
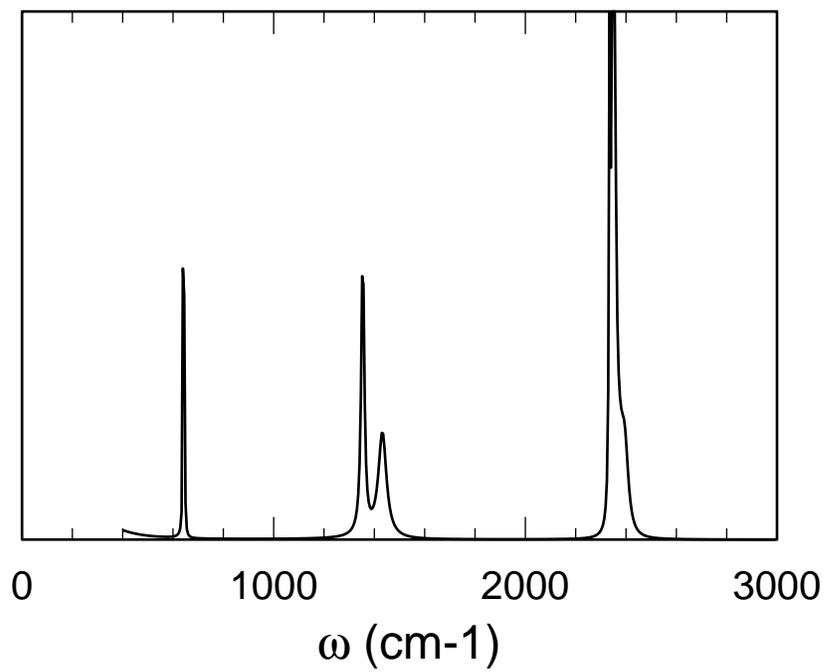
\end{document}